\documentclass[sigconf]{acmart}

\AtBeginDocument{%
  \providecommand\BibTeX{{%
    \normalfont B\kern-0.5em{\scshape i\kern-0.25em b}\kern-0.8em\TeX}}}

\copyrightyear{2020}
\acmYear{2020}
\setcopyright{rightsretained}
\acmConference[IUI '20 Workshops]{March 17, 2020}{Cagliari, Italy}

\acmDOI{}

\begin{document}

\title{Business (mis)Use Cases of Generative AI}









\author{Stephanie Houde, Vera Liao, Jacquelyn Martino, Michael Muller}
\author{David Piorkowski, John Richards, Justin Weisz, Yunfeng Zhang}
\email{Stephanie.Houde@ibm.com, Vera.Liao@ibm.com, jmartino@us.ibm.com, michael_muller@us.ibm.com}
\email{David.Piorkowski@ibm.com, ajtr@us.ibm.com, jweisz@us.ibm.com, zhangyun@us.ibm.com}
\affiliation{IBM Research}

\renewcommand{\shortauthors}{Houde et al.}

\begin{abstract}
Generative AI is a class of machine learning technology that learns to generate new data from training data. While deep fakes and
media-and art-related generative AI breakthroughs have recently
caught people's attention and imagination, the overall area is in its infancy for business use. Further, little is known about generative AI's potential for malicious misuse at large scale. Using co-creation design fictions with AI engineers, we explore the plausibility and severity of business misuse cases.
\end{abstract}

\ccsdesc[500]{Human-centered computing}
\ccsdesc[300]{Social and professional topics~Computing / technology policy}
\ccsdesc[300]{Applied computing~Law, social and behavioral sciences}

\keywords{Generative AI, design fiction, human-in-the-loop}


\maketitle

\section{Introduction}
As we think about a future where humans and AI partner in creative activities, we consider how generative models \cite{tsourikov2017generative, van2013generative, zittrain2005generative, radford2019language, 2015arXiv151106434R} could impact current businesses and possibly create new ones.  While deep fakes and media-and art-related generative AI breakthroughs have recently caught people's attention and imagination \cite{agarwal2019limits, guera2018deepfake, korshunov2019vulnerability, li2018exposing, shtefaniukdeepfake}, the area overall is in its infancy for business use. In this paper, we take an inverse approach to business cases and propose business ``misuse'' cases from the ``bad actor'' point of view. Using a practice called design fictions \cite{ blythe2018imaginary, dunne2001design, dunne2013speculative, sterling2009design, tanenbaum2014design, tanenbaum2016limits}, we engaged with software engineers who are expert in AI technologies. We provided three half-page fictions about possible harmful applications of generative technologies as probes for use in our co-creation exercises. As we guided the engineers' exploration of the three fabricated misuse business cases, we learned their points of view on factors such as scenario plausibility, seriousness, and prevention. 

The contributions of this paper include:

\begin{itemize}
\item participant-created future scenarios based on generative AI capabilities in text, audio, and video
\item reactions to possible generative AI design fiction scenarios \item discussion responses to specific probes on scenario plausibility, seriousness, ways things might be worse, ways things might be better, and prevention
\end{itemize}

The remainder of this paper is organized as follows: Section 2 discusses related work in AI and Human Centered Data Science (HCDS), and then introduces design fictions as a method for ``prototyping'' and discussing possible future outcomes of AI. Section 3 describes how we used design fictions with experts in AI software engineering to explore potential business misuse cases. Section 4 presents our results. Section 5 offers concluding thoughts.

\section{Related Work}
Strong claims are made about the promises and current successes of AI and data science \cite{durrant2015data, hey2009fourth, miller2017future, van2014data}. While some of these claims are projected for the future \cite{ dobre2014intelligent}, Agarwal and Dhar editorialized in 2014 that ``This is powerful... we are in principle already there'' \cite{agarwal2014big}. Meanwhile to the dystopian extreme, scholars warn about the ``mythology'' of working with big data, that the quantitative nature of data gives a false illusion that all data-driven outcomes are objective, ethical, or true \cite{boyd2012critical}. Increasingly, scholars in the emerging field of Human Centered Data Science (HCDS) \cite{aragon2016developing, boukhelifa2019exploratory, muller2019human, wang2019human} have begun to investigate data science practices, showing the necessary, responsible, and increasingly accountable human activities that take place between data and models \cite{boukhelifa2019exploratory, feinberg2017design, gitelman2013raw, kandel2011wrangler, muller2019data, passi2017data, passi2018trust, pine2015politics, rattenbury2017principles}.

In this paper we extend this emerging work by examining possible applications of recent developments in generative AI methods. Rather than waiting to ``see what happens,'' we apply the low-cost method of design fictions to prototype \cite{bleecker2009design, dunne2001design, dunne2013speculative, lindley2015researching, lindley2014machine, pargman2017sustainability, skirpan2018ad, sondergaard2016periodshare, sterling2009design, tanenbaum2014design, tanenbaum2016limits, wakkary2015material, wong2016product} possible future applications. Rather than depending on our own views, we engage with the thoughtful and creative contributions of knowledgeable colleagues \cite{baumer2014chi, blythe2014chatbots, blythe2018imaginary, cheon2018futuristic, light2011democratising, muller2017exploring, wong2017eliciting}.

There has been increasing interest in HCI of projecting the future of technologies through design fictions (DFs) -- fictional scenarios in the form of narratives, concepts, prototypes, enactments, and games. \cite{blythe2014chatbots, blythe2018imaginary, blythe2015solutionism, cheon2018futuristic}. DFs have been applied to the design of new technology \cite{cila2017products}, to explore how future users may adopt a technology \cite{cheon2018futuristic}, and as critical tools to anticipate the social and political consequences of technologies \cite{ dourish2014resistance, tanenbaum2014design}.

Design fictions as a methodology spring from several sources. While many cite Sterling's introductory definition of ``deliberate use of diegetic prototypes to suspend disbelief about change'' \cite{sterling2009design}, relevant forms have also been explored in the fields of Participatory Design \cite{beeson2000dialogue, bodker1991design}, and Future Studies \cite{pargman2017sustainability}. A predecessor to DFs in HCI research is design scenarios \cite{wright2014ethical} - simple vignettes to illustrate the use of technologies to be developed. DFs sometimes focus on the design of concrete ``near future'' technologies \cite{ bleecker2009design}, constructing a discursive fictional space for speculating about the effects of such technologies in the future. In particular, DFs has been embraced for revealing values associated with new technologies \cite{muller2017exploring, tanenbaum2014design}.

There are many forms and usages of design fiction. We differentiate between research works that create DFs as the end products (e.g., to frame new design concepts that do not yet exist \cite{brown2016ikea, heibeck2014sensory}), or as critical tools \cite{blythe2014chatbots, blythe2018imaginary}), and those using DFs as probes -- ``critical narratives to elicit open-ended responses from potential future users [stakeholders] of proposed technologies'' \cite{schulte2016homes}. 

Our use of DFs falls in the latter category. Some also refer to it as participatory design fiction, in contrast to previous approaches where DFs are created only by researchers \cite{muller2017exploring}. For example, Shulte et al. described a 5-step method to create design fictions on the topic of smart houses, and illustrated how the method can be used for research and design purposes. Recently, several works explored using the Story Completion Method (SCM) in design fiction. SCM was first introduced in psychotherapy and qualitative research in psychology \cite{shah2018therapists}. Wood et al. used SCM for speculative stories to explore the future vision of Virtual Reality pornography \cite{wood2017they}. Cheon and Su introduced Futuristic Autobiographies \cite{cheon2018futuristic} as a way to elicit perspectives on the future of technologies and conducted a case study on the future of robotics. The method starts by posing stories involving the participant as a character in a future state, as grounded in background research work, and invites the participant to complete the autobiography.

\section{Methodology}
Our exploratory study into potential business misuse cases began with the creation of design fictions focusing on the use of generative AI for malicious purposes. We wrote three scenarios, each representing a different generative media type: text, audio, or video. Then, using these scenarios as a springboard, we led co-creation sessions with AI experts to gain their views on possible worst-case business scenarios. We conducted these sessions in counterbalanced order across participants, to control for novelty, fatigue, and other possible order effects.

\subsection{Interviews}
We conducted co-creation sessions with six participants (2 female, 4 male) from our research organization. All participants are software engineers practiced in AI techniques. After a level-setting description of generative techniques as a class of machine learning and their potential use for malicious intent, we presented our potential future scenarios. Each scenario was presented in two parts. The first part is a design fiction from the near future (2020). After reading the first part, we asked each participant to speculate on a worst case misuse of generative AI in ten more years (2030) and invent their own future scenario. After hearing their thoughts, we shared the second part which was our version of the future in 2030. With both our version and their version in mind, we asked them to respond to the following questions:
    
\begin{enumerate}
\item Is the scenario plausible?
\item Is this a serious problem?
\item Can you make the scenario worse?
\item Can you make the scenario better?
\item Can anything be done to prevent this?
\item Is there a way out of this?
\item Who is the right person, organization, or entity to improve it?
\end{enumerate}

Participants spent about 15 minutes reading and discussing each scenario for a total time of about 45 minutes.
  
\subsection{Scenarios}

\subsubsection{Scenario A. Instant Author Story}
\hfill

\textit{Part 1.} It's 2020 (and you've just seen this popup ad): Hey! I bet you've tried to write the Great American Novel several times but you can't seem to get past the first chapter. We get it. Like you, we find original writing to be really difficult and basically unrewarding. Luckily, we've created MakeMyStory.com that takes someone else's content, does a deep analysis of it, and emits it in a form that is as compelling as the original work but different enough that no one will say it's not yours. How cool is that? As an added bonus, your new work can be emitted in any language, increasing the number of countries where it can be sold.

\textit{Part 2.} It's 2030 now and Amazon is completely flooded with works by previously unknown writers. A few prominent authors have taken MakeMyStory to court to try to have the web site shut down. Others have sued the nominal authors of these works but have not been successful. This is due to a few things. First, MakeMyStory has declared their software to be proprietary so it can't be examined and the courts have gone along with that so far (go figure). Second, the work that is claimed to be derived passes every known test for originality. (Interesting side note: perhaps an investigation into how the original authors determined that their work had been stolen might lead to new techniques for automatically finding such ``adaptations''. ) Third, a recently-added REMIX capability allows multiple original works to be ``blended'' making derivation tracking almost impossible. So ... check back in in another couple of years, assuming anyone is still bothering to read by then.

\subsubsection{Scenario B. Fake Smoking Evidence Insurance Claim Story}
\hfill

\textit{Part 1.} It's 2020 and Jane just accepted a new job. As part of her onboarding process, she is answering some health questions for insurance purposes, including, of course, whether she smokes. That one is simple...never. Later that year, Jane, develops shortness of breath, confirmed to be the early stages of emphysema. Confident that her health insurance will cover the costs she begins some very expensive treatment. When the first batch of bills is rejected by the insurance company she calls to find out why and is simply shocked. Here is the conversation:

\textit{Representative}: I'm sorry Jane, but based on your medical history, you will not be covered for this procedure. Your history of smoking disqualifies you from coverage.

\textit{Jane:} What are you talking about, I've never smoked in my life!

\textit{Representative:} Well, I'm looking at some videos online that beg to differ. Did you attend a wedding last month?

\textit{Jane:} Yes, it was my best friend's wedding. But I don't understand... what are you talking about?

\textit{Representative:} You were wearing a blue dress, correct?

\textit{Jane:} Yes.

\textit{Representative:} Well ma'am, I'm watching a video from the wedding where you are very clearly smoking a cigarette. As such your request for coverage has been denied.

Jane knows that video had to have been faked, but she has no idea how to prove it. Let's just hope her savings hold out.

\textit{Part 2.} In what seems to be a pattern of corruption, insurance companies are denying medical claims due to video ``evidence'' of risky behavior ranging from base jumping to travel to active war zones. Car insurance claims are being rejected due to ``evidence'' of preexisting damage. Even life insurance claims are being denied based on ``evidence'' that the deceased is still alive. Legislation was passed to make these practices illegal, but without the ability to detect when a video is faked, and without the ability to truly prove the legitimacy of a piece of digital content, insurance companies are able to continue these practices to fatten their bottom line.

\subsubsection{Scenario C. Audio Evidence Story}
\hfill

\textit{Part 1.} It's 2020, and the following story just appeared on page 14 of the New York Times: Expert testimony for the defense was presented today in the case of Walter Milgrim, a prominent New York investor accused of arranging the contract murder of his wife. In earlier testimony, the prosecution's expert witness had played an audio recording claimed to have been captured by a passenger on the Chappaqua train platform on the night of the murder. In it, a person sounding like Milgrim and a person sounding like the trigger man could be heard discussing the means of payment. While the voices were somewhat muffled, the recording seemed authentic. This was cleverly refuted by the defense expert, who commissioned a separate recording at the train station. In that recording, a low hum was heard due to wind passing through the overhead wires. This hum was not present in the prosecution's recording, but it was to be expected due to windy conditions that night. Reasonable doubt? We'll see.

\textit{Part 2.} It's 2030 and a number of criminal convictions have been thrown into turmoil by the discovery of an audio tool that can synthesize ultra-realistic ambient soundscapes based on thousands of parameters. For example, all manner of weather conditions can be simulated including how those conditions interact with physical features of a location. Seemingly-authentic crowd noises (from the faint shuffle of feet on snow to the sounds of people almost having to shout over traffic or being nearly drowned out by the arrival of a train or airplane) add to the realism. A whistleblower has alleged that a number of high profile cases in the Southern District of New York relied on recordings generated by this tool. It is unknown how many cases have been corrupted and there seems to be no way to determine which recordings were real and which were, in fact, generated.

\section{Results}
Two members of the research team independently reviewed the transcripts and audio recordings of each of the three scenarios in each of the six sessions, looking for key themes and comments. These were then assembled into a combined data set. As no clear differences between the three scenarios were found, we present observations summarized by themes within session phases, starting with participant comments during their invented 2030 futures, followed by their reactions to our envisioned 2030 futures, and ending with their responses to the focused questions.

\subsection{Participant Invented Futures}
\subsubsection{Ubiquity} Participants anticipated a significant increase in fakes due to reduced costs. Many of these fakes would likely be created by the general public. It was noted that large scale automated faking might be particularly dangerous since it would spread adverse impact beyond what a few bad actors would likely do. Interestingly, the very ubiquity of fakes might also reduce the impact of any particular fake since less weight would be given to any single piece of content. Widespread surveillance, with verifiable provenance, might become more 
important as a means of countering faked evidence, and privacy laws might 
require revisions as policy-makers learn more about how vulnerable groups could be affected by the proliferation of fakes, and in what ways.
Not surprisingly, there will be increased need for reliable detection of fakes.
\subsubsection{Quality} 
When informants were discussing \textit{their own participatory fictions}, they
suggested that machine generated text may actually be better than human authored text, at least for some purposes. It was also noted that faked content may actually be acceptable if it is used for entertainment rather than as evidence.
\subsubsection{Ownership} There was a belief that all work was, in some sense, derived from what came before. This view led to the thought that automatically derived content was perhaps acceptable.
\subsubsection{Markets} Niche markets for hand-crafted content might emerge with ``written by humans'' sections being tucked in the back of book stores and online sales sites. A greatly expanded need for digital forensics in cases of disputed ownership and other matters of evidence might cause this to become an emerging business opportunity.
\subsubsection{Technology} It was felt that models may be able to detect features of fakes that would not be apparent to humans.

\subsection{Reactions To Our Envisioned Futures}
\subsubsection{Ubiquity} People might start creating verified videos of themselves to protect against fakes. This is already happening with always-on dash cams in both private and police cars. Because of the ease of fake creation, multiple independent sources of evidence might become even more required.
\subsubsection{Quality} 
When informants were discussing \textit{our probe fictions}, they
thought that there was a real possibility of a descent into mediocrity as fakes fed on themselves in a ``never ending resounding echo chamber of cyber-generated nonsense.'' Several thought we risked a loss of ``art,'' both in the creation of and the appreciation of content, with creatives having significantly reduced value. It is possible this would not be sustained, though, as much generated content might be ``boring as hell.''
\subsubsection{Ownership} One participant wondered if AI would have to be acknowledged as a co-creator of content.
\subsubsection{Markets} Creating content for a single consumer would be feasible, both for entertainment and education. This would upend the current mass distribution market model.
\subsubsection{Technology} It was thought that there would be a significant ramp up of research in detecting machine-generated content. Progress here would likely be aided by the fact that detection capability would increase as the corpus of fakes for training grew.
On the other hand, widespread watermarking and the use of block chains might obviate the need for much of this detection capability as we could rely on immutable metadata to prove originality.

\subsection{Answers To Focused Questions}
\subsubsection{Plausibility} There were mixed views on the plausibility of 2030 scenarios. Some felt that similar futures were inevitable due to reduced costs of fakes and often malevolent human nature. Others considered these futures unlikely due to legal, regulatory, and market forces, the emergence of technical detection and provenance mechanisms, and the belief that societies have generally been good at authenticating things they care about (e.g., money, passports).
\subsubsection{Seriousness} There were also mixed views on the seriousness of these possible futures; if a future was considered implausible for the reasons noted above it was also viewed as not serious. On the other hand, it was mentioned that a single blemish in a person's health record could profoundly impact the ability to obtain insurance coverage. Thus, quite serious outcomes might arise from even isolated incidents of faking.

\subsubsection{How It Could Be Worse} One participant felt that generative models interactively posing as humans might be particularly harmful. We already see low-tech variants of this with, for example, people posing as somewhat-distant relatives needing money to get them out of difficulty during foreign travel. The synthesis of realistic human speech was considered harder than the synthesis of ambient audio, but the potential impact could be much worse. On a different note, the coherent faking of multiple sources might further complicate  use as evidence. One possible reaction to all this would be governments banning the use of generative AI. If \textit{only} governments were allowed to use it, the results could be quite dire. Finally, if these sorts of futures come to pass, people might simply stop caring about truth altogether.
\subsubsection{How It Could Be Better} Several potentially beneficial uses of generative AI were offered. In one, the AI could generate realistic time-aged portraits of lost loved ones to convey the sense that they were still present in peoples' lives. In another, high-quality generated speech could be used by those with severe communication disabilities. Related to this, 
a generative AI might align the sounds of diverse human accents 
in an audio stream for improved mutual intelligibility when time was critical (e.g., in an emergency call). Finally, the ability to cheaply produce content could provide value to currently under-served markets.
\subsubsection{How It Could Be Prevented} Answers to the ``Prevention'' and ``Way Out'' questions were quite similar and are combined here. Most felt that some combination of technology, regulation, and legislation could prevent the worst of these future scenarios. Data traceability and auditability, ubiquitous detection of fake content and watermarking of captured content in personal devices, and new legislation that finally caught up with emerging threats were all mentioned. Education and training were noted by many participants as important; making people aware of the threats posed by fake content and equipping them with skills to detect it might diminish the viral quality of fakes.
\subsubsection{Who Might Improve} Here too, some combination of technology providers, regulators (especially international organizations), and legislative bodies were viewed as likely contributors. Specialized oversight mechanisms might also arise to meet the needs of academic research and publishing, medical research, art competitions, and so on. One participant thought that AI itself might assist in doing legal research and crafting good law to counter threats. Finally, consumers and critics might rate and review content for quality and originality in a manner not too dissimilar from today.

\section{Concluding Thoughts}
Our participants' views were both thoughtful and highly varied. Even though their comments stayed at a quite general level, they sometimes came to contrasting conclusions when discussing their own future scenarios as opposed to the future scenarios that we created (particularly regarding quality issues). These types of contrasts in perspectives have been important in other projects with participatory forms of design fictions \cite{baumer2014chi, blythe2014chatbots, blythe2018imaginary, cheon2018futuristic, light2011democratising, muller2017exploring, wong2017eliciting}. Taken together, these contrasting results suggest that \textit{including informants in collaborative interpretive spaces} can be a powerful method for increasing the knowledge of participants and researchers alike.

Some informants thought scenarios of the sort we sketched for 2030 were inevitable while others thought they were unlikely. Drawing on their own expertise, their thoughts on how technology might counter possible threats were more nuanced than their thoughts about regulation and legislation. Perhaps the most interesting dynamics arose in connection with what many saw as an ``arms race'' between generation and detection capabilities. There was optimism that the coming ubiquity of fakes will lead, in a manner not unlike a natural ecosystem, to an abundance of training data for high-quality detectors. That, along with progress in digital provenance and updated regulations and legislation, may prevent the worst of these future scenarios from coming to pass. It will be important to expand these HCDS approaches \cite{aragon2016developing, boukhelifa2019exploratory, muller2019human, wang2019human} to include more diverse informants in the future, for a more societally-grounded vision and critique of the implications of generative AI.

\bibliographystyle{ACM-Reference-Format}
\bibliography{bibliography}

\end{document}